\pdfoutput=1
\documentclass[11pt]{article}

\usepackage[final]{acl}

\usepackage{times}
\usepackage{latexsym}

\usepackage[T1]{fontenc}
\usepackage[utf8]{inputenc}

\usepackage{microtype}

\usepackage{inconsolata}

\usepackage{graphicx}

\usepackage{colortbl}
\usepackage{booktabs}
\usepackage{multirow}
\usepackage{cleveref}
\usepackage{tcolorbox}
\usepackage{enumitem}
\usepackage{caption}
\usepackage{subcaption}
\usepackage[frozencache,cachedir=minted-cache]{minted}

\crefname{table}{table}{tables}
\Crefname{table}{Table}{Tables}

\crefname{figure}{figure}{figures}
\Crefname{figure}{Figure}{Figures}

\title{Self-Constructed Context Decompilation with Fined-grained Alignment Enhancement}

\author{
Yunlong Feng, Dechuan Teng, Yang Xu, Honglin Mu, \\ 
\textbf{Xiao Xu, Libo Qin, Qingfu Zhu\thanks{Corresponding Author}, Wanxiang Che} \\ 
Harbin Institute of Technology, China \\
\{ylfeng,dcteng,yxu,hlmu,xxu,lbqin,qfzhu,car\}@ir.hit.edu.cn
}

\begin{document}
\maketitle
\begin{abstract}
  Decompilation transforms compiled code back into a high-level programming language for analysis when source code is unavailable.
  Previous work has primarily focused on enhancing decompilation performance by increasing the scale of model parameters or training data for pre-training.
  Based on the characteristics of the decompilation task, we propose two methods: (1) Without fine-tuning, the Self-Constructed Context Decompilation (sc$^2$dec) method recompiles the LLM's decompilation results to construct pairs for in-context learning, helping the model improve decompilation performance.
  (2) Fine-grained Alignment Enhancement (FAE), which meticulously aligns assembly code with source code at the statement level by leveraging debugging information, is employed during the fine-tuning phase to achieve further improvements in decompilation.
  By integrating these two methods, we achieved a Re-Executability performance improvement of approximately 3.90\% on the Decompile-Eval benchmark, establishing a new state-of-the-art performance of 52.41\%.
  The code, data, and models are available at \href{https://github.com/AlongWY/sccdec}{https://github.com/AlongWY/sccdec}.
\end{abstract}

\section{Introduction}

\label{sec:intro}

\begin{figure}[t]
  \centering
  \includegraphics[width=0.7\linewidth,page=1]{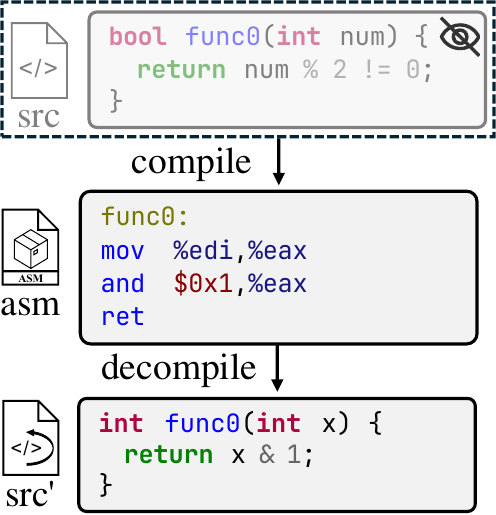}
  \caption{Pipeline of the decompilation. The input for decompilation tasks is typically assembly, with the source code invisible. Additionally, the source code obtained through decompilation usually does not exactly match the original code.}
  \label{fig:overview}
\end{figure}

Decompilation is the process of converting compiled machine code or bytecode back into a high-level programming language. This process is typically used to analyze how the software works when the source code is not accessible~\cite{decompilation1, decompilation2_rnn, btc, lmpa, slade, nova, refine_decompile}. Many tools have been developed for decompilation, like Ghidra~\cite{ghidra} and IDA Pro~\cite{idapro}.
However, these tools often struggle to generate human-readable code.
The main challenge in decompilation is that it's hard to fully reconstruct the source code, especially details such as variable names~\cite{variable_name} and primary structures~\cite{for_loop}, which are frequently lost during the compilation process.

Recent advances in large language models (LLMs) have prompted researchers to view programming languages as distinct linguistic systems, utilizing pre-trained code LLMs to accomplish various coding tasks~\cite{starcoder, codellama, deepseekcoder, gpt4}.
These models have demonstrated significant performance improvements over traditional techniques~\cite{codelm_study1, codelm_study2}, making it feasible to apply LLMs to decompilation challenges.
For instance, Transformer-based models such as Slade~\cite{slade} and Nova~\cite{nova} have shown potential in using language models to translate binary code back into more readable and structured source code.
Recently, \citet{llm4decompile} developed and released the first open-source LLM focused on decompilation, alongside the construction of the first decompilation benchmark to evaluate the decompilation capabilities of models.
These efforts primarily treat the decompilation task as a translation task, training a robust decompilation model on extensive synthetic data \cite{btc, slade, nova, llm4decompile}.

By analyzing the characteristics of decompilation tasks, we have found two features:
\begin{itemize}[noitemsep]
  \item The decompiled code generated by the model is usually compilable.
  \item The compiler can produce external debugging information for the debugger.
\end{itemize}
Based on the characteristics of the decompilation task, we propose two approaches:
(1) \textit{Self-Constructed Context Decompilation} (sc$^2$dec), a tuning-free approach, leverages the decompilation results from the model and specific compiler to derive ground-truth $(assembly\ code, source\ code)$ pair closely related to the test sample.
By incorporating the obtained pair as a demonstration example within the context, we can fully leverage the model's in-context learning capabilities to enhance performance.
(2) \textit{Fine-grained Alignment Enhancement} (FAE) introduces a new step-by-step decompilation training objective in addition to the end-to-end decompilation training objective that relies on large-scale pre-training for implicit alignment.
By leveraging debugging information designed for debuggers, we can align assembly code with high-level code at a finer statement level.
This allows the model to more faithfully preserve the original functionality after decompilation.

Combining these two methods, we have achieved an approximately 3.90\% improvement in re-executability performance based on llm4decompile model~\cite{llm4decompile}, reaching a new state-of-the-art performance of 52.41\% re-executability.

To summarize, our contributions are as follows:

\begin{itemize}
  \item We introduce \textit{Self-Constructed Context Decompilation}, which leverages the compilability of decompilation results to construct better contexts.
  \item We introduce \textit{Fine-grained Alignment Enhancement}, which fine-tunes the model using fine-grained alignment data between assembly code and source code derived from debug information. We also propose how to synthesize its training data automatically.
  \item The experiments on the Decompile-Eval benchmark show that our proposed model achieves new state-of-the-art results.
\end{itemize}

\section{Related Works}
\label{sec:rel}

Decompilation has a long history, centered on converting binary files into human-readable source code. Traditional tools such as Hex-Rays IDA Pro~\cite{idapro} and Ghidra~\cite{ghidra} predominantly rely on analyzing a program's control flow graph (CFG) to perform this task~\cite{decompilation1}. These tools operate by scrutinizing instructions within assembly code to construct CFGs and identify common programming structures like if-else statements and loops~\cite{for_loop}. However, they rely on intricate and error-prone expert-built rule systems that frequently fail with optimized binary code, a common practice in commercial software development. Moreover, traditional tools often generate outputs that closely resemble assembly code representations, translating variables into registers and utilizing low-level operations such as goto statements. This output is not only challenging to comprehend but also occasionally cannot be recompiled~\cite{ioacc2}.

Inspired by neural machine translation, researchers have begun to conceptualize decompilation as a translation task, wherein machine-level instructions are converted into readable source code using neural networks. Initial endeavors employing recurrent neural networks (RNNs) demonstrated limited success~\cite{decompilation2_rnn}. However, recent advancements in natural language processing (NLP) have facilitated the application of pre-trained language models (LMs) in decompilation. Notable examples include BTC~\cite{btc}, Slade~\cite{slade}, and Nova~\cite{nova}, which mark significant strides in the field.
Recently, \citet{llm4decompile} created and released the first open-source large language models specifically for decompilation and established an evaluation benchmark that considers the re-compilability and re-executability of decompiled code.

\section{Method}
\label{sec:method}

\begin{figure}[t]
  \centering
  \includegraphics[page=2,width=\linewidth]{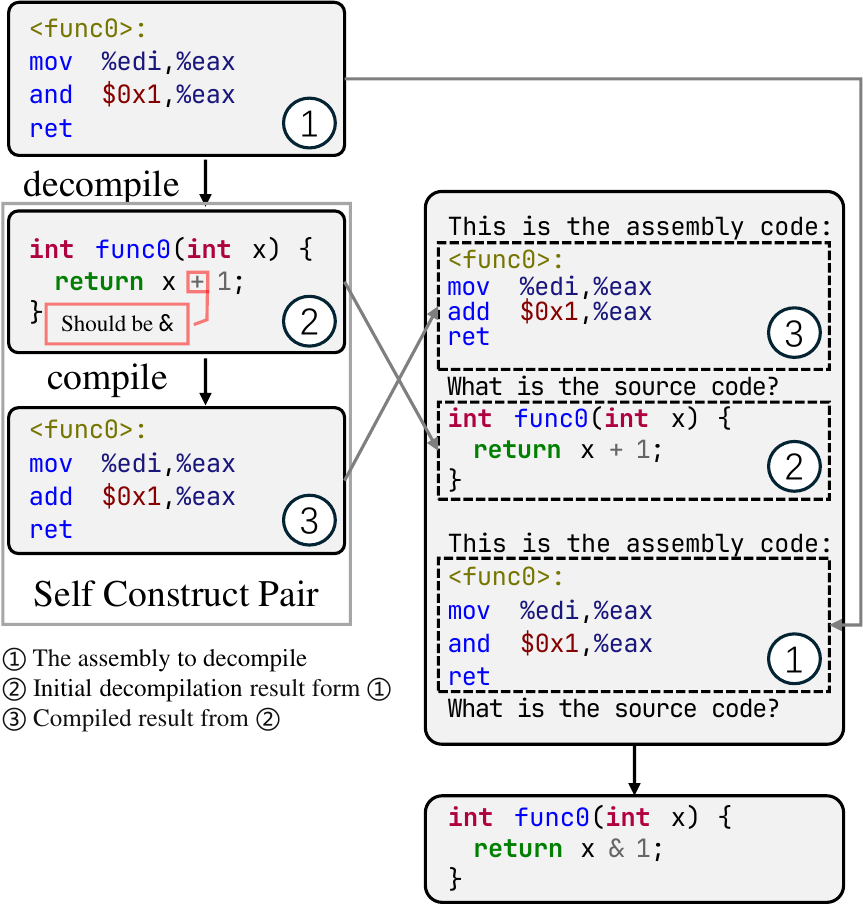}
  \caption{The pipeline for Self-Constructed Context Decompilation operates as follows: when the LLM decompiles and generates compilable code, we compile this code to construct the context, and then use it to decompile the initial assembly code again.}
  \label{fig:sccdec}
\end{figure}

In this section, we introduce the Self-Constructed Context Decompilation method and Fine-grained Alignment Enhancement fine-tuning.
\Cref{fig:sccdec,fig:stepdec} illustrate the methods.

\subsection{Self-Constructed Context Decompilation}
\label{sec:method1}

In this section, we introduce the \textbf{S}elf-\textbf{C}onstructed \textbf{C}ontext \textbf{Dec}ompilation (sc$^2$dec) method designed to enhance the performance of decompilation models.
The method recompiles the decompiled code generated by the model to construct the context.
The sc$^2$dec method involves the following key steps, as the \Cref{fig:sccdec} shows:

\begin{enumerate}[noitemsep]
  \item \textbf{Decompile}: Initially, the model decompiles the assembly code back to the target programming language. We call it the initial decompilation result.
  \item \textbf{Compile and Disassembly}: Then we recompile the initial decompilation result to obtain the corresponding assembly code if the generated code is compilable.
  \item \textbf{Constructed Context}: The self-constructed context is then formed by concatenating the generated C code with its corresponding assembly code.
  \item \textbf{Decompile Again}: Decompile the assembly code with the self-constructed context again.
\end{enumerate}

In this context, the assembly code precedes the C code, resembling an example in the context. The final input to the model is a concatenation of the self-constructed context (with assembly code first and C code second) and the original assembly code. This method leverages the additional context provided by the generated C code and its corresponding assembly code to improve the overall decompilation performance.

\subsection{Fine-grained Alignment Enhancement}
\label{sec:method2}

\begin{figure}[t]
  \centering
  \includegraphics[width=0.9\linewidth,page=3]{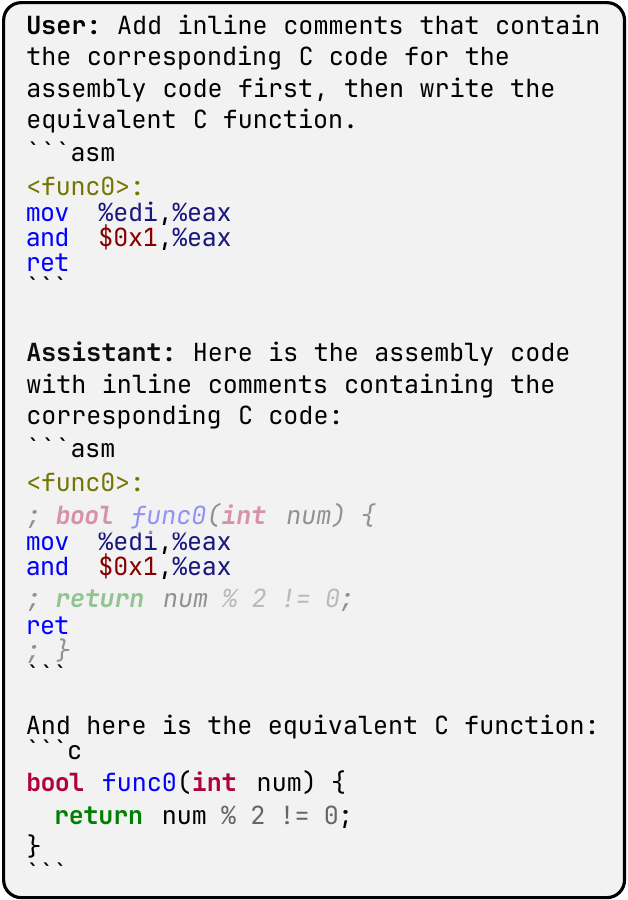}
  \caption{An example of step-by-step decompilation is presented, where the training objective requires the model to generate C code progressively after each assembly block. Fine-tuning the model with this objective aids in learning the fine-grained correspondences between assembly and C code.}
  \label{fig:stepdec}
\end{figure}

To enhance the model's ability to perceive the correspondence between assembly code and its corresponding source code, we propose a Fine-grained Alignment Enhancement (FAE) method, which constructs fine-grained alignment data between assembly code and source code derived from debug information. The method primarily includes two training objectives: the End-to-end Decompilation Objective and the Step-by-Step Decompilation Objective.

\begin{itemize}[noitemsep]
  \item \textbf{End-to-end Decompilation Objective}: In the end-to-end decompilation task, the model takes assembly code as input and directly generates the corresponding decompiled C code. The goal of this task is to enable the model to translate from assembly language to C language without any intermediate steps. This objective aims to prevent catastrophic forgetting in the model. During training, we increase the context length and ensure that the same sample is not truncated, which enhances the model's ability to align assembly and C code.

  \item \textbf{Step-by-step Decompilation Objective}: As the \Cref{fig:stepdec} shows, in the step-by-step decompilation process, an intermediate form containing both the assembly code and the corresponding C code is first generated, followed by the generation of the complete C code. This step-by-step approach allows the model to better understand the correspondence between assembly code and C code. Training the model with the step-by-step decompilation objective can enhance the model's ability to finely align assembly code with C code. It is important to note that this form is used only during training and not during inference.
\end{itemize}

During the training process, we employed two different training objectives to enhance the model's decompile performance: an end-to-end decompile objective and a step-by-step decompile objective. They all take assembly code as input, but their output formats are different.

\subsubsection{Data Processing}

By following the steps below, we can synthesize some fine-grained alignment training data. The synthesized data, as shown in \Cref{fig:stepdec}, is in a format of interleaved assembly and code output. We expect the model to learn finer-grained alignment between assembly and code from such data.

\begin{enumerate}[noitemsep]
  \item \textbf{Compilation}:
        We employed the ``gcc'' compiler to compile the selected function code into binary shared libraries at various optimization levels (including O0, O1, O2, and O3). The debug option was enabled during compilation to ensure the inclusion of DWARF debugging information in the compiled output. This debugging information is crucial for subsequent disassembly and parsing steps.
  \item \textbf{Disassembly}:
        Subsequently, we used the ``objdump'' tool to disassemble the compiled binary shared libraries. The ``\texttt{-S}'' option was utilized to include source code in the disassembly output. Additionally, we used the ``\texttt{--source-comment=;}'' option to prepend each line of source code with a ``\texttt{;}'' symbol, facilitating easier parsing in later stages.
  \item \textbf{Data Parsing}:
        The disassembly process yielded a mixed output containing both assembly code and source code, organized in an alternating sequence of source code followed by assembly code. We parsed the output to extract the corresponding assembly and source code for each function, ensuring accurate alignment between the two.
  \item \textbf{Data Reorganization}:
        Finally, we reorganized the parsed content into a format where the assembly code precedes the corresponding source code in an alternating sequence.
\end{enumerate}

Through these steps, we constructed a high-quality training dataset that ensures each sample contains comprehensive source code information along with the corresponding assembly code. This dataset will be used to train our model, aiming to enhance its performance in tasks such as code understanding and compilation optimization.

\section{Experiments}
\label{sec:exp}

\subsection{Evaluation}

In this section, we will introduce the benchmark used in our evaluation process, Decompile-Eval \cite{llm4decompile}, which is specifically designed to assess the decompilation capabilities of large language models.

The Decompile-Eval benchmark is adapted from the HumanEval benchmark, which includes 164 problems initially designed for code generation tasks. These problems were translated into the C programming language, and the corresponding assembly code was generated at four optimization levels (O0, O1, O2, and O3). The correctness of the decompilation results was tested using the test cases from HumanEval. The primary metrics of the Decompile-Eval benchmark are as follows:

\begin{itemize}[noitemsep]
  \item \textbf{Re-compilability}: This metric evaluates whether the decompiled code produced by our model can be successfully recompiled into executable binary without errors. A high recompilability rate indicates that the decompiled code is syntactically correct and adheres to the constraints of the target language (in this case, C).
  \item \textbf{Re-executability}: This metric assesses the functional correctness of the decompiled code. Specifically, it measures whether the recompiled binaries produce the expected outputs when executed. The correctness of the output is determined using the testing methodology provided by the HumanEval dataset, ensuring a comprehensive evaluation of the logical accuracy of the decompiled code.
\end{itemize}

In our experiment, we pay more attention to the Re-executability, as it more accurately reflects the overall decompilation capability of the model.

\subsection{Training Details}

\paragraph{Implementation} We utilize Lora \cite{hu2022lora} to fine-tune the llm4decompile-6.7b obtained on Hugging Face \cite{huggingface}., with rank set to 32, alpha set to 64 and target set to all projection layers \footnote{q\_proj, k\_proj, v\_proj, o\_proj, gate\_proj, up\_proj, down\_proj}. The optimizer is AdamW \cite{adamw}, with a learning rate of 5e-5. The maximum sequence length is set to 16384, and the learning rate scheduler type is cosine, with a warm-up period of 20 steps. The training process was conducted for one epoch. LlamaFactory \cite{llamafactory} and FlashAttention 2 \cite{flashattention2} were used for the fine-tuning of the model. All experiments were done with an A100-SXM4-80GB. We use greedy decoding for all experiments.

\paragraph{Training data} We selected 10,000 samples from the train\_real\_compilable subset of Exebench \cite{exebench} to synthesis the training data. The selected functions exclusively utilize the standard C library and do not include additional data structures. The training data were synthesized with gcc 11.4 provided by Ubuntu 22.04.

\subsection{Baselines}

We present several baselines in our experiments and demonstrate the effectiveness of our proposed methods.

\subsubsection{Models}
\begin{itemize}[noitemsep]
  \item \textbf{GPT 4}: One of the most powerful OpenAI models, known for its advanced language understanding and generation capabilities \cite{gpt4}.
  \item \textbf{deepseek-chat}: A powerful conversational AI model from DeepSeek, excelling in generating coherent and contextually relevant dialogues \cite{deepseekv2}.
  \item \textbf{llm4decompile-6.7b}: The 6.7B open-access decompilation LLM pre-trained on 15 billion tokens of C source code and the corresponding assembly code \cite{llm4decompile}\footnote{We use the latest version, \href{https://huggingface.co/LLM4Binary/llm4decompile-6.7b-v1.5}{llm4decompile-6.7b-v1.5}.\label{footnote:llm4decompile}}.
  \item \textbf{llm4decompile-6.7b+FAE}: The new model was obtained by applying the Fine-grained Alignment Enhancement method for further fine-tuning based on the llm4decompile-6.7b\textsuperscript{ \ref{footnote:llm4decompile}}.
\end{itemize}

\subsubsection{Methods}

\begin{figure}[t]
  \centering
  \includegraphics[width=\linewidth,page=4]{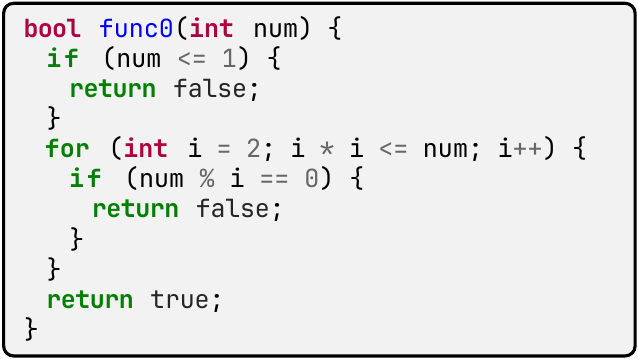}
  \caption{The example for 1-shot learning. In this example, we have tried to cover common control logic such as if statements, loops, and early returns as much as possible. The example will be compiled with the same optimization level as the target assembly code.}
  \label{fig:one-shot}
\end{figure}

\begin{table*}[t]
  \centering
  \resizebox{\textwidth}{!}{%
    \begin{tabular}{lrrrrrrrrrr}
      \toprule
      \multicolumn{1}{c}{\multirow{2}{*}{Method}} & \multicolumn{5}{c}{Re-Compilability} & \multicolumn{5}{c}{Re-Executability}                                                                                                                                         \\ \cmidrule(l){2-11}
      \multicolumn{1}{c}{}                         & O0                                   & O1                                   & O2             & O3             & AVG            & O0             & O1             & O2             & O3             & AVG            \\ \hline
      \rowcolor[rgb]{0.93,0.93,0.93}\multicolumn{11}{c}{\textbf{GPT-4 \cite{gpt4}}}                                                                                                                                                                                      \\
      vanilla                                      & 76.83                                & 67.68                                & 67.07          & 57.93          & 67.38          & 20.73          & 13.41          & 13.41          & 10.98          & 14.63          \\
      \multicolumn{1}{r}{~~~~+sc$^{2}$dec}         & 76.22                                & 67.03                                & 65.24          & 56.10          & 66.15          & 32.93          & 20.12          & 19.51          & 13.41          & 21.49          \\
      1-shot                                       & 76.22                                & 78.05                                & 73.78          & 70.73          & 74.70          & 31.71          & 20.12          & 19.51          & 21.34          & 23.17          \\
      \multicolumn{1}{r}{+sc$^{2}$dec}             & 74.39                                & 76.22                                & 71.95          & 68.90          & 72.87          & 40.24          & 25.61          & 23.78          & 23.17          & 28.20          \\
      \rowcolor[rgb]{0.93,0.93,0.93}\multicolumn{11}{c}{\textbf{Deepseek Chat \cite{deepseekv2} }}                                                                                                                                                                       \\
      vanilla                                      & 62.20                                & 51.83                                & 47.56          & 45.73          & 51.83          & 10.37          & 5.49           & 4.88           & 5.49           & 6.55           \\
      \multicolumn{1}{r}{+sc$^{2}$dec}             & 57.93                                & 50.00                                & 46.95          & 45.12          & 50.00          & 13.41          & 6.71           & 6.71           & 7.32           & 8.54           \\
      1-shot                                       & 67.68                                & 70.73                                & 64.63          & 56.71          & 64.94          & 13.41          & 10.37          & 7.93           & 7.32           & 9.76           \\
      \multicolumn{1}{r}{+sc$^{2}$dec}             & 65.85                                & 70.12                                & 62.20          & 53.66          & 62.96          & 17.68          & 11.59          & 10.37          & 9.15           & 12.20          \\ \hline \hline
      \rowcolor[rgb]{0.93,0.93,0.93}\multicolumn{11}{c}{\textbf{llm4decompile-6.7b \cite{llm4decompile}}}                                                                                                                                                                \\
      vanilla                                      & 92.80                                & 93.05                                & 90.73          & \textbf{94.02} & 92.65          & 70.24          & 45.49          & 40.61          & 38.05          & 48.60          \\
      \multicolumn{1}{r}{+sc$^{2}$dec}             & 92.68                                & 92.80                                & 90.12          & 93.41          & 92.26          & 71.34          & \textbf{47.56} & 45.98          & \textbf{41.71} & \textbf{51.65} \\
      1-shot                                       & 93.05                                & \textbf{93.29}                       & \textbf{90.98} & 93.90          & \textbf{92.80} & 70.37          & 44.88          & 41.46          & 37.68          & 48.60          \\
      \multicolumn{1}{r}{+sc$^{2}$dec}             & 93.05                                & 92.68                                & 90.37          & 93.29          & 92.35          & \textbf{71.59} & 47.20          & \textbf{46.34} & 41.46          & \textbf{51.65} \\
      retrieval                                    & \textbf{94.02}                       & 88.05                                & 84.51          & 85.12          & 87.93          & 65.73          & 32.68          & 38.66          & 36.22          & 43.32          \\
      \rowcolor[rgb]{0.93,0.93,0.93}\multicolumn{11}{c}{\textbf{llm4decompile-6.7b + FAE}}                                                                                                                                                                               \\
      vanilla                                      & \textbf{92.68}                       & \textbf{92.44}                       & \textbf{93.66} & \textbf{93.17} & \textbf{92.99} & 67.80          & 47.68          & 45.73          & 42.32          & 50.88          \\
      \multicolumn{1}{r}{+sc$^{2}$dec}             & 92.07                                & 91.59                                & 91.71          & 92.32          & 91.92          & \textbf{70.24} & \textbf{48.54} & \textbf{47.56} & \textbf{43.29} & \textbf{52.41} \\
      1-shot                                       & \textbf{92.68}                       & 92.32                                & \textbf{93.66} & 92.80          & 92.87          & 67.68          & 47.20          & 45.61          & 41.95          & 50.61          \\
      \multicolumn{1}{r}{+sc$^{2}$dec}             & 92.07                                & 91.71                                & 91.46          & 91.95          & 91.80          & \textbf{70.24} & 48.05          & 47.20          & 43.17          & 52.16          \\ \bottomrule
    \end{tabular}%
  }
  \caption{The main results of different methods across four optimization levels (O0, O1, O2, O3), as well as their average scores (AVG). The results in bold represent the optimal performance, while those underlined indicate the second-best performance. More results can be found in the \Cref{tab:extra-result}.}
  \label{tab:main-result}
\end{table*}

\begin{itemize}[noitemsep]
  \item \textbf{vanilla}: Using specific prompts for different models to directly request decompilation.

        \begin{tcolorbox}[title = {Prompt For GPT-4 and Deepseek Chat}]
          \begin{minted}[breaklines]{text}
What is the C source code of the assembly code below:
```asm
<asm>
```
\end{minted}
        \end{tcolorbox}

        \begin{tcolorbox}[title = {Prompt For llm4decompile}]
          \begin{minted}[breaklines]{text}
# This is the assembly code: 
<asm>

# What is the source code?
\end{minted}
        \end{tcolorbox}
  \item \textbf{1-shot}: We write a fixed example as the context \cite{gpt3}. The example will be compiled with the same optimization level as the target assembly code. The example shown in \Cref{fig:one-shot}. In this example, we have tried to cover common control logic such as if statements, loops, and early returns as much as possible.
  \item \textbf{retrieval}: We use the retrieval method (BM25 \cite{bm25s}) to get the context. The context is generated by searching for the most similar assembly code in the training set.
  \item \textbf{sc$^{2}$dec}: The self-constructed context decompilation we described in the \Cref{sec:method1}.
  \item \textbf{1-shot+sc$^{2}$dec}: We apply the sc$^{2}$dec on the 1-shot method. Note that the 1-shot example can only be applied to generate the initial decompilation result in \Cref{fig:sccdec}.
\end{itemize}

\begin{table*}[t]
  \centering
  \resizebox{\textwidth}{!}{%
    \begin{tabular}{lrrrrrrrrrr}
      \toprule
      \multicolumn{1}{c}{\multirow{2}{*}{Method}} & \multicolumn{5}{c}{Re-Compilability} & \multicolumn{5}{c}{Re-Executability}                                                                                                                                         \\ \cmidrule(l){2-11}
                                                    & O0                                   & O1                                   & O2             & O3             & AVG            & O0             & O1             & O2             & O3             & AVG            \\ \midrule
      sc$^{2}$dec                                   & 92.07                                & 91.59                                & 91.71          & 92.32          & 91.92          & \textbf{70.24} & \textbf{48.54} & 47.56          & 43.29          & \textbf{52.41} \\
      sc$^{2}$dec (O0)                              & \textbf{92.68}                       & \textbf{92.20}                       & 91.71          & 92.32          & 92.23          & \textbf{70.24} & 48.90          & 47.68          & 42.93          & 51.83          \\
      sc$^{2}$dec (O1)                              & \textbf{92.68}                       & 91.59                                & \textbf{93.54} & \textbf{92.93} & \textbf{92.68} & 68.41          & \textbf{48.54} & 47.44          & 43.05          & 51.86          \\
      sc$^{2}$dec (O2)                              & \textbf{92.68}                       & \textbf{92.20}                       & 91.71          & 92.32          & 92.23          & 67.80          & 48.90          & 47.68 & 42.93          & 51.68          \\
      sc$^{2}$dec (O3)                              & \textbf{92.68}                       & 91.59                                & 92.93          & 92.32          & 92.38          & 67.80          & 47.07          & \textbf{47.80}          & \textbf{43.41} & 51.52          \\ \bottomrule
    \end{tabular}%
  }
  \caption{The results of our method (llm4decompile-6.7b + FAE) across optimization levels. For cases such as ``Ubuntu gcc 11.4 (O0)'', we assume that the optimization level of asm code is entirely unknown and select a specific optimization level to compile the code to construct the context.}
  \label{tab:optimization}
\end{table*}
\begin{figure}[t]
  \centering
  \includegraphics[width=\linewidth,page=5]{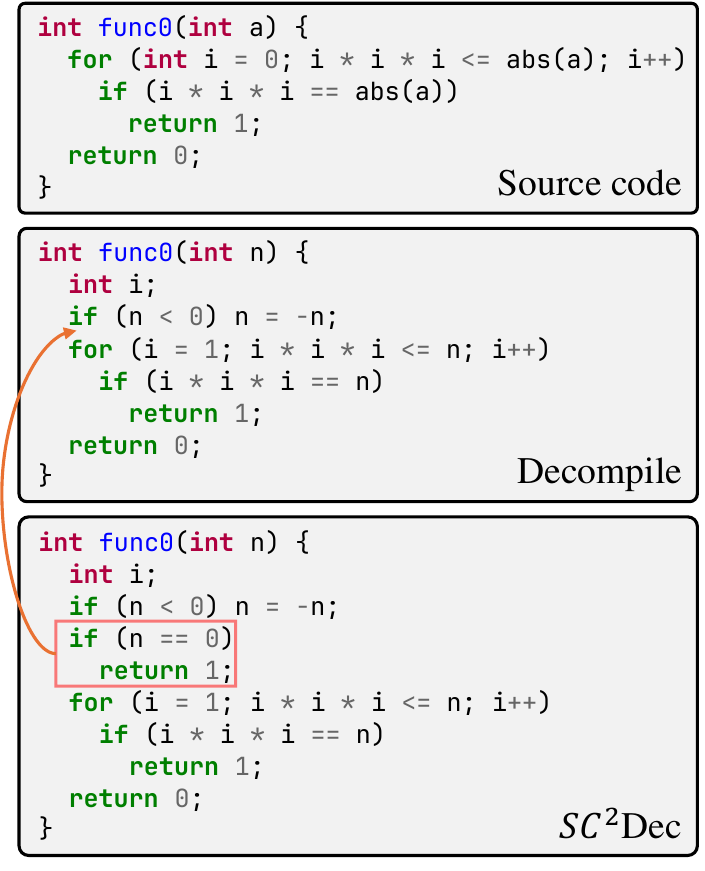}
  \caption{A case study for sc$^2$dec, which is based on the llm4decompile-6.7b with FAE. By applying the sc$^2$dec method, the model detects and fixes the mismatch between the code and assembly.}
  \label{fig:case-study}
\end{figure}

\subsection{Results}

In this section, we present and analyze the performance of different models. Our experiments aim to evaluate the impact of various techniques, including one-shot, self-constructed context decompilation, and fine-tuning, on the model's performance. The main results are shown in \Cref{tab:main-result}.

\paragraph{Performance of the Base Model}
Through a comparative analysis of the experimental results in \Cref{tab:main-result}, it is evident that different models exhibit significant differences in re-executability across various optimization levels (O0, O1, O2, O3). The base model, llm4decompile-6.7b, performs best at the O0 optimization level with a score of 70.24\%, but its performance gradually declines at other optimization levels, with an average score of 48.60\%. Among models not fine-tuned with decompilation tasks, GPT-4 achieved the best performance, with an average re-executability score of 14.64\%. Generally, their results on re-compilability metrics are significantly higher than their performance on re-executability metrics.

\paragraph{Context Sensitivity} By comparing vanilla, 1-shot, and retrieval methods based on the llm4decompile-6.7b, it is observed that the model's performance can even decline after providing a sample. In contrast, the dynamic samples generated through our self-constructed context method yield an improvement in executability of over 2\% before and after fine-tuning. This indicates that inappropriate samples may even have a detrimental effect in the decompilation domain, while our method achieves a nearly stable performance enhancement without the need for retrieval.

\paragraph{1-shot is a Strong Baseline}
As shown in \Cref{tab:main-result}, utilizing the 1-shot method significantly enhances the model's performance.
Specifically, for GPT-4, the model's re-executability performance increased from 14.64\% to 23.17\%, an approximate improvement of 7\%.
In the open-source model Deepseek Chat, performance improved from 6.55\% to 9.76\%, reflecting a 3.21\% enhancement.
However, in the case of the llm4decompile model, which had already been trained on decompilation tasks, there was no performance improvement.
Moreover, in the llm4decompile model fine-tuned with Fine-grained Alignment Enhancement, performance actually decreased from 50.88\% to 50.61\%, a reduction of approximately 0.27\%.

\paragraph{Sc²dec Brings Further Performance Improvement}
The \Cref{tab:main-result} shows that applying sc²dec independently results in performance improvements of 6.86\% and 1.99\% on GPT-4 and Deepseek Chat, respectively.
These improvements are lower than those achieved by the 1-shot method, likely due to the model's low Re-Compilability metric.
In other words, a substantial portion of the code obtained through decompilation is un-compilable, which impedes the application of the sc²dec method.
Furthermore, when combined with the 1-shot method, sc²dec can further enhance performance, achieving an additional 5\% improvement for GPT-4, bringing it to 28.20\%, and a 2.44\% improvement for Deepseek Chat, bringing it to 12.20\%.
In the case of llm4decompile-6.7b, due to the models' relatively high Re-Compilability metric, the sc²dec method significantly outperformed the 1-shot method, increasing performance from 48.60\% to 51.65\% and from 50.88\% to 52.41\%.

\begin{table*}[t]
  \resizebox{\textwidth}{!}{%
    \begin{tabular}{lrrrrrrrrrr}
      \toprule
      \multicolumn{1}{c}{\multirow{2}{*}{Model}} & \multicolumn{5}{c}{Re-Compilability} & \multicolumn{5}{c}{Re-Executability}                                                                                                                                         \\ \cmidrule(l){2-11}
      \multicolumn{1}{c}{}                       & O0                                   & O1                                   & O2             & O3             & AVG            & O0             & O1             & O2             & O3             & AVG            \\ \hline
      w/o tune                                   & 92.80                                & \textbf{93.05}                       & 90.73          & 94.02          & 92.65          & 70.24          & 45.49          & 40.61          & 38.05          & 48.60          \\
      \rowcolor[rgb]{0.93,0.93,0.93}\multicolumn{11}{c}{FAE (End-to-End \& Step-by-Step Decompilation Objective)}                                                                                                                                                      \\
      vanilla                                    & 92.68                                & 92.44                                & \textbf{93.66} & \textbf{93.17} & \textbf{92.99} & 67.80          & 47.68          & 45.73          & 42.32          & 50.88          \\
      \multicolumn{1}{r}{~~~~~~+sc$^{2}$dec}           & 92.07                                & 91.59                                & 91.71          & 92.32          & 91.92          & 70.24          & \textbf{48.54} & \textbf{47.56} & \textbf{43.29} & \textbf{52.41} \\
      1-shot                                     & 92.68                                & 92.32                                & \textbf{93.66} & 92.80          & 92.87          & 67.68          & 47.20          & 45.61          & 41.95          & 50.61          \\
      \multicolumn{1}{r}{+sc$^{2}$dec}           & 92.07                                & 91.71                                & 91.46          & 91.95          & 91.80          & 70.24          & 48.05          & 47.20          & 43.17          & 52.16          \\
      \rowcolor[rgb]{0.93,0.93,0.93}\multicolumn{11}{c}{End-to-End Decompilation Objective}                                                                                                                                                                            \\
      vanilla                                    & \textbf{93.9}                        & 90.61                                & 90.61          & 91.34          & 91.62          & 69.51          & 43.29          & 42.2           & 40.73          & 48.93          \\
      \multicolumn{1}{r}{+sc$^{2}$dec}           & 92.8                                 & 90.24                                & 90.00          & 90.12          & 90.79          & \textbf{71.34} & 46.10          & 43.78          & 42.44          & 50.91          \\
      1-shot                                     & \textbf{93.9}                        & 90.73                                & 90.37          & 91.22          & 91.55          & 69.51          & 44.02          & 42.68          & 40.61          & 49.21          \\
      \multicolumn{1}{r}{+sc$^{2}$dec}           & 92.8                                 & 90.24                                & 89.76          & 90.00          & 90.7           & 71.22          & 46.46          & 44.27          & 42.56          & 51.13          \\ \bottomrule
    \end{tabular}%
  }
  \caption{The results of ablation study. All results are based on the llm4decompile-6.7b model.}
  \label{tab:tuning}
\end{table*}

\paragraph{Best Performance with Combined Methods}
As the \Cref{tab:main-result} shows, by combining our Step-by-Step Decompile tasks and context-based decompilation methods, we ultimately achieved approximately a 3.90\% performance improvement, surpassing the performance of these methods when applied independently.
This demonstrates that fine-tuning and self-construct context methods are orthogonal.
Most notably, the combined method of fined-grained alignment
enhancement and self-construct context decompilation achieves the highest scores across all optimization levels expect ``O0'', with scores of 48.54\%, 47.56\%, and 43.29\% at the O1, O2, and O3 levels, respectively, and an overall average score of 52.41\%.

\subsection{Analysis}

In this section, we will further analyze the experimental results to study the effectiveness and sensitivity of our method:

\paragraph{Mismatched Optimization Levels Leads to Performance Degradation}
As the \Cref{tab:optimization} shows, the performance can degrade by 1\% to 2\% overall when constructing contexts that mismatch the optimization levels of the target assembly code.
However, specific optimization levels may exhibit different behaviors.
For instance, using a fixed O3 optimization level to construct contexts while decompiling assembly optimized with O2 does not result in significant performance degradation and even better. \textit{\textbf{In practical scenarios, using O3 to construct contexts for binaries with unknown optimization levels might be a reasonable choice}}.

\paragraph{Case study of sc$^2$dec} As the \Cref{fig:case-study} shows, the compiler has expanded the abs function, reducing the overhead of function calls. The model initially generated a piece of compilable but incorrect code during direct decompilation, omitting the case where the variable $n$ equals $0$.
We compiled and disassembled this code to construct a new sample.
By applying the sc$^2$dec method with the context, the model successfully identified the missing part in the short segment of assembly code and generated the correct code, thus rectifying the error from the previous inference.

\paragraph{Contribution of FAE} The results in \Cref{tab:main-result} show that the application of Fine-grained Alignment Enhancement (FAE) for further fine-tuning the model significantly enhanced its re-executability performance across all optimization levels except ``O0''.
Notably, at optimization levels O2 and O3, the scores improved by 5.12\% and 4.27\%, reaching 45.73\% and 42.32\%, respectively. The average performance increased by 2.28\%, achieving 50.88\%.
This demonstrates the effectiveness of the Fine-grained Alignment Enhancement method.

\paragraph{Ablation Study of FAE} The results of ablation experiments in \Cref{tab:tuning} have demonstrated the effectiveness of our proposed Step-by-Step Decompile training objective.
Removing the training objective results in approximately a 2\% decline in Re-Executability performance, regardless of whether the sc$^{2}$dec method is used, and notably, a 3.78\% decline at the ``O2'' optimization level.
However, the performance of the 1-shot remains unaffected, supporting our hypothesis that the mismatch between the training data form and the inference form leads to performance degradation.
Furthermore, retaining only the End-to-end Decompilation training objective still yields little performance improvement, about 0.33\%.
This improvement can be attributed to the longer context provided during training, as the samples are not truncated during training, allowing the model to fully observe the assembly code and its corresponding C code.

\section{Conclusion}
\label{sec:conclusion}

In this paper, we introduce two methods based on the characteristics of the decompilation task: Self-Constructed Context Decompilation, which leverages the compilability of decompilation results to construct better contexts, and Fine-grained Alignment Enhancement, which fine-tunes the model using fine-grained alignment data between assembly code and source code derived from debug information.
By integrating these methods, we achieved a 3.90\% improvement in decompilation performance, reaching a new state-of-the-art level of 52.41\%.

\section*{Acknowledge}

We gratefully acknowledge the support of the National Natural Science Foundation of China (NSFC) via grant 62236004, 62206078, 62441603 and 62476073.

\section*{Limitations}

The sc$^2$dec depends on the decompiled code generated by the model. If the decompiled code cannot be recompiled, sc$^2$dec will not be able to benefit from it, necessitating that the decompiled code has high re-compilability performance. For Fine-grained Alignment Enhancement, we created a Step-by-Step Decompile training objective, training with only 10,000 samples across four optimization levels (O0-O3), without validating its performance on larger datasets or bigger models. Moreover, fine-tuning led to a further decline in performance in the 1-shot scenario.

\section*{Potential risks}
Decompilation is a technique that converts compiled binary code back into human-readable source code. While decompilation can be legal and useful in certain contexts, it also entails several potential risks. The primary potential risks are as follows:

\begin{itemize}
  \item Intellectual Property Infringement: Decompilation may violate the copyright and licensing agreements of software. Unauthorized decompilation can lead to copyright infringement, thereby instigating legal disputes.
  \item Security Risks: Decompiled code may expose the internal structure and implementation details of the software, providing attack vectors for hackers. Malicious actors can exploit this information to identify and exploit vulnerabilities in the software.
  \item Ethical Concerns: Decompiling and analyzing another's code can be considered unethical, especially when done without authorization.
\end{itemize}

\bibliography{custom}

\appendix

\onecolumn

\section{Training examples}

\begin{figure}[h]
  \centering
  \includegraphics[width=\linewidth,page=6]{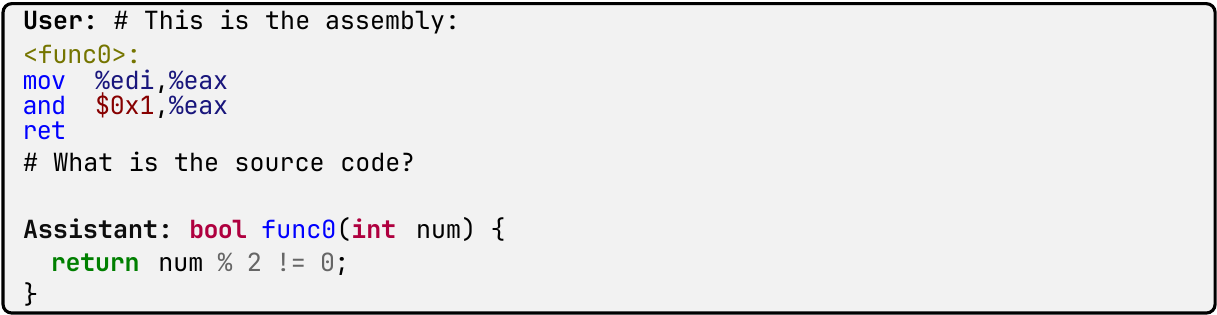}
  \caption{The training example of end-to-end decompile objective.}
  \label{fig:end2end}
\end{figure}

\begin{figure}[h]
  \centering
  \includegraphics[width=\linewidth,page=7]{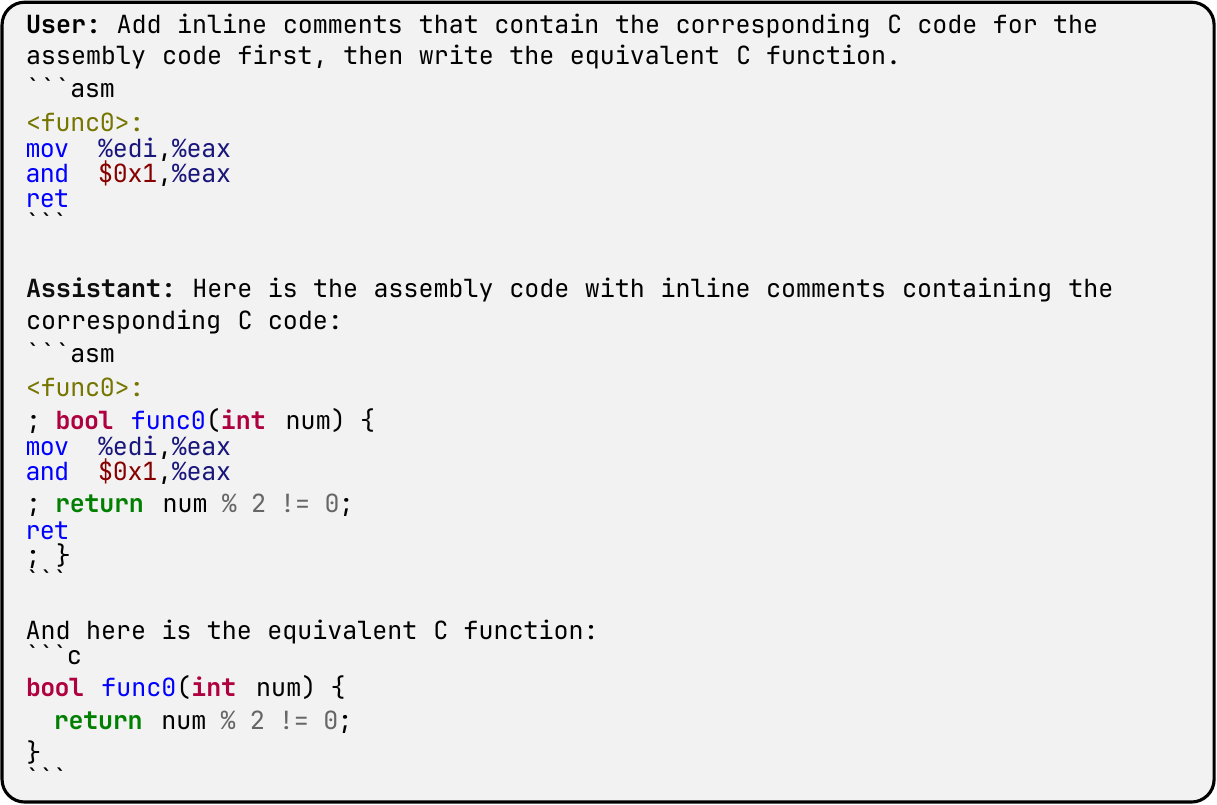}
  \caption{The training example of step-by-step decompile objective. Note that the workflow is only for training.}
  \label{fig:stepbystep}
\end{figure}

\section{Extra Results}

\begin{table*}[t]
  \centering
  \resizebox{0.9\textwidth}{!}{%
    \begin{tabular}{lrrrrrrrrrr}
      \toprule
      \multicolumn{1}{c}{\multirow{2}{*}{Compiler}} & \multicolumn{5}{c}{Re-Compilability} & \multicolumn{5}{c}{Re-Executability}                                                                 \\ \cmidrule(l){2-11}
                                                     & O0                                   & O1                                   & O2    & O3    & AVG   & O0    & O1    & O2    & O3    & AVG   \\ \midrule
      x86-64 gcc 10.5                                & 92.07                                & 90.98                                & 91.71 & 92.32 & 91.77 & 70.85 & 47.32 & 47.80 & 43.05 & 52.26 \\
      x86-64 gcc 11.4                                & 92.07                                & 91.59                                & 92.32 & 92.32 & 92.07 & 70.85 & 48.66 & 47.68 & 42.68 & 52.47 \\
      x86-64 gcc 12.3                                & 92.07                                & 91.59                                & 92.32 & 92.32 & 92.07 & 70.85 & 47.68 & 46.95 & 42.56 & 52.01 \\
      x86-64 gcc 13.3                                & 92.07                                & 91.59                                & 92.32 & 92.32 & 92.07 & 70.85 & 47.93 & 46.59 & 42.68 & 52.01 \\
      x86-64 gcc 14.1                                & 92.07                                & 91.83                                & 92.44 & 92.56 & 92.23 & 70.85 & 47.07 & 45.73 & 41.95 & 51.40 \\
      x86-64 clang 14.0.0                            & 92.68                                & 92.44                                & 93.66 & 93.17 & 92.99 & 69.02 & 48.05 & 45.73 & 42.44 & 51.31 \\
      x86-64 clang 15.0.0                            & 92.68                                & 92.44                                & 93.66 & 93.17 & 92.99 & 69.02 & 48.29 & 45.73 & 42.32 & 51.34 \\
      x86-64 clang 16.0.0                            & 92.68                                & 92.2                                 & 93.66 & 93.17 & 92.93 & 69.02 & 48.05 & 46.46 & 42.32 & 51.46 \\
      x86-64 clang 17.0.1                            & 92.68                                & 92.44                                & 93.66 & 93.17 & 92.99 & 69.02 & 48.05 & 46.71 & 42.32 & 51.52 \\
      x86-64 clang 18.1.0                            & 92.68                                & 92.44                                & 93.66 & 93.17 & 92.99 & 69.02 & 48.17 & 46.46 & 42.32 & 51.49 \\ \bottomrule
    \end{tabular}%
  }
  \caption{The re-executability of differnet models across different compilers. We selected the popular three versions of both the clang and gcc series in \href{https://godbolt.org}{Compiler Explorer}. The compiler of Compiler Explorer may differ from the compilers provided by Ubuntu in some default compilation options, which might result in slight differences in the generated assembly code.}
  \label{tab:compiler}
\end{table*}

\paragraph{The Model is More Sensitive to Compiler Series than to Versions:} As the \Cref{tab:compiler} shows, using a compiler different from the one used during training to build the context can lead to significant performance degradation. The model performs better when the context is built using the GCC compiler series, whereas the performance is slightly worse when using the clang series. This may be due to the fact that the context used during both the continued pre-training and fine-tuning stages was constructed with the GCC compiler series. When different versions of GCC are used for construction, the performance of sc$^2$dec stabilizes around 52\% for the fine-tuned model except ``x86-64 gcc 14.1''. In contrast, when the context is built using the clang series, the performance of sc$^2$dec is slightly lower and more consistent, stabilizing around 51\% for the fine-tuned model.

\begin{table*}[t]
  \centering
  \resizebox{0.9\textwidth}{!}{%
    \begin{tabular}{lrrrrrrrrrr}
      \toprule
      \multicolumn{1}{c}{\multirow{2}{*}{Method}} & \multicolumn{5}{c}{Re-Compilability} & \multicolumn{5}{c}{Re-Executability}                                                                 \\ \cmidrule(l){2-11}
      \multicolumn{1}{c}{}                         & O0                                   & O1                                   & O2    & O3    & AVG   & O0    & O1    & O2    & O3    & AVG   \\ \hline
      \rowcolor[rgb]{0.93,0.93,0.93}\multicolumn{11}{c}{\textbf{GPT-4 \cite{gpt4}}}                                                                                                              \\
      vanilla                                      & 76.83                                & 67.68                                & 67.07 & 57.93 & 67.38 & 20.73 & 13.41 & 13.41 & 10.98 & 14.63 \\
      \multicolumn{1}{r}{~~~~+sc$^{2}$dec}         & 76.22                                & 67.03                                & 65.24 & 56.10 & 66.15 & 32.93 & 20.12 & 19.51 & 13.41 & 21.49 \\
      1-shot                                       & 76.22                                & 78.05                                & 73.78 & 70.73 & 74.70 & 31.71 & 20.12 & 19.51 & 21.34 & 23.17 \\
      \multicolumn{1}{r}{+sc$^{2}$dec}             & 74.39                                & 76.22                                & 71.95 & 68.90 & 72.87 & 40.24 & 25.61 & 23.78 & 23.17 & 28.20 \\
      \rowcolor[rgb]{0.93,0.93,0.93}\multicolumn{11}{c}{\textbf{GPT-3.5 Turbo \cite{chatgpt}}}                                                                                                   \\
      vanilla                                      & 20.73                                & 25.00                                & 29.27 & 21.95 & 24.24 & 4.27  & 3.66  & 3.05  & 3.05  & 3.51  \\
      \multicolumn{1}{r}{+sc$^{2}$dec}             & 18.29                                & 22.56                                & 28.66 & 20.12 & 22.41 & 5.49  & 3.66  & 3.66  & 4.27  & 4.27  \\
      1-shot                                       & 49.39                                & 46.34                                & 42.07 & 40.85 & 44.66 & 10.37 & 7.32  & 6.10  & 5.49  & 7.31  \\
      \multicolumn{1}{r}{+sc$^{2}$dec}             & 42.07                                & 42.68                                & 36.59 & 34.15 & 38.87 & 12.20 & 9.15  & 6.71  & 7.93  & 8.99  \\
      \rowcolor[rgb]{0.93,0.93,0.93}\multicolumn{11}{c}{\textbf{Deepseek Chat \cite{deepseekv2} }}                                                                                               \\
      vanilla                                      & 62.20                                & 51.83                                & 47.56 & 45.73 & 51.83 & 10.37 & 5.49  & 4.88  & 5.49  & 6.55  \\
      \multicolumn{1}{r}{+sc$^{2}$dec}             & 57.93                                & 50.00                                & 46.95 & 45.12 & 50.00 & 13.41 & 6.71  & 6.71  & 7.32  & 8.54  \\
      1-shot                                       & 67.68                                & 70.73                                & 64.63 & 56.71 & 64.94 & 13.41 & 10.37 & 7.93  & 7.32  & 9.76  \\
      \multicolumn{1}{r}{+sc$^{2}$dec}             & 65.85                                & 70.12                                & 62.20 & 53.66 & 62.96 & 17.68 & 11.59 & 10.37 & 9.15  & 12.20 \\
      \rowcolor[rgb]{0.93,0.93,0.93}\multicolumn{11}{c}{\textbf{Deepseek Coder \cite{deepseekcoder}}}                                                                                            \\
      vanilla                                      & 66.46                                & 57.93                                & 47.56 & 51.83 & 55.95 & 9.15  & 6.10  & 6.71  & 6.98  & 7.01  \\
      \multicolumn{1}{r}{+sc$^{2}$dec}             & 62.80                                & 54.27                                & 43.29 & 50.00 & 52.59 & 12.20 & 7.93  & 9.15  & 8.54  & 9.45  \\
      1-shot                                       & 57.93                                & 53.66                                & 52.44 & 48.17 & 53.05 & 11.59 & 10.37 & 8.54  & 7.93  & 9.60  \\
      \multicolumn{1}{r}{+sc$^{2}$dec}             & 56.10                                & 51.83                                & 51.22 & 46.34 & 51.37 & 15.24 & 11.58 & 10.37 & 8.54  & 11.43 \\ % \midrule
      \rowcolor[rgb]{0.93,0.93,0.93}\multicolumn{11}{c}{\textbf{llm4decompile-6.7b \cite{llm4decompile}}}                                                                                        \\
      vanilla                                      & 92.80                                & 93.05                                & 90.73 & 94.02 & 92.65 & 70.24 & 45.49 & 40.61 & 38.05 & 48.60 \\
      \multicolumn{1}{r}{+sc$^{2}$dec}             & 92.68                                & 92.80                                & 90.12 & 93.41 & 92.26 & 71.34 & 47.56 & 45.98 & 41.71 & 51.65 \\
      1-shot                                       & 93.05                                & 93.29                                & 90.98 & 93.90 & 92.80 & 70.37 & 44.88 & 41.46 & 37.68 & 48.60 \\
      \multicolumn{1}{r}{+sc$^{2}$dec}             & 93.05                                & 92.68                                & 90.37 & 93.29 & 92.35 & 71.59 & 47.20 & 46.34 & 41.46 & 51.65 \\
      \rowcolor[rgb]{0.93,0.93,0.93}\multicolumn{11}{c}{\textbf{llm4decompile-6.7b + FAE}}                                                                                                       \\
      vanilla                                      & 92.68                                & 92.44                                & 93.66 & 93.17 & 92.99 & 67.80 & 47.68 & 45.73 & 42.32 & 50.88 \\
      \multicolumn{1}{r}{+sc$^{2}$dec}             & 92.07                                & 91.59                                & 91.71 & 92.32 & 91.92 & 70.24 & 48.54 & 47.56 & 43.29 & 52.41 \\
      1-shot                                       & 92.68                                & 92.32                                & 93.66 & 92.80 & 92.87 & 67.68 & 47.20 & 45.61 & 41.95 & 50.61 \\
      \multicolumn{1}{r}{+sc$^{2}$dec}             & 92.07                                & 91.71                                & 91.46 & 91.95 & 91.80 & 70.24 & 48.05 & 47.20 & 43.17 & 52.16 \\ \bottomrule
    \end{tabular}%
  }
  \caption{The extra results of different methods across four optimization levels (O0, O1, O2, O3), as well as their average scores (AVG). The results in bold represent the optimal performance, while those underlined indicate the second-best performance.}
  \label{tab:extra-result}
\end{table*}

\end{document}